# Oxygen Reduction Reaction on Platinum Nanocatalysts Produces Long-Lived, Hysteretic Oxygenated Adsorbates

Jaehyeon Kim[1,2], Lalith Krishna Samanth Bonagiri[1,2], Fujia Zhao[1,2], and Yingjie Zhang[1,2,3]*

1. Department of Materials Science and Engineering, University of Illinois, Urbana, Illinois 61801, United States

2. Materials Research Laboratory, University of Illinois, Urbana, Illinois 61801, United States

3. Beckman Institute for Advanced Science and Technology, University of Illinois, Urbana, Illinois 61801, United States

*Correspondence to: yjz@illinois.edu

**ABSTRACT:** Aqueous electrocatalysis generates oxygenated intermediates at catalyst surfaces. While intermediate species on single-crystal catalysts have been observed, the nature and evolution of surface oxygenated species on industrially relevant nanoparticle (NP) catalysts remain largely unknown. Here, using *in situ* Raman spectroscopy, we tracked the formation and potential-dependent evolution of oxygenated adsorbates in alkaline media on NP catalysts with an active platinum (Pt) surface. By comparing spectroscopic features in Ar- vs $O_2$-saturated electrolytes, we determined three key intermediates produced by the oxygen reduction reaction (ORR): adsorbed OOH, OH, and $O_2$. In contrast to the conventional wisdom that intermediates exist only during catalytic reactions, we found these oxygenated adsorbates to be highly long-lived and hysteretic, and to persist even after the termination of ORR. This adsorbate-retention effect exhibits a modest dependence on the surface oxidation state and the electrolyte cations ($K^+$ vs $Li^+$), and is likely facilitated by the heterogeneous nature of the catalyst surface. The results highlight the complexity of surface adsorption structures on realistic catalysts, which often extends beyond that captured by measurements or simulations on model single-crystal surfaces.

## INTRODUCTION

Aqueous electrocatalytic reactions, such as oxygen reduction (ORR),[1–3] oxygen evolution (OER),[4,5] and $CO_2$ reduction reaction ($CO_2$RR),[6,7] form oxygenated adsorbates on catalyst surfaces. These species covalently bind to the catalyst surface[8] and can modulate catalytic activities by altering adsorption energies,[1,9] the density of catalytically active sites,[10,11] catalyst structure,[12] and/or reactant transport to the surface.[8] As the reaction proceeds, the coverage and composition of these adsorbates evolve, and their impact on activity can change accordingly. Therefore, understanding how surface adsorption structure and coverage evolve under reactive conditions is essential for establishing mechanistic links between surface structure and catalytic activity.

ORR is a central process in fuel cells[8] and broadly relevant to metal–air batteries[13] and biological energy conversion such as aerobic respiration.[14,15] Alkaline fuel cells (AFCs) and alkaline anion-exchange membrane fuel cells (AAEMFCs) continue to attract interest due to their reduced corrosiveness and potential economic advantages compared to proton-exchange membrane fuel cells.[16,17] However, alkaline ORR remains less explored, and existing mechanistic studies of alkaline ORR have focused on model systems such as single crystals.[2,8,18–20] While these approaches provide valuable atomistic insights by simplifying the surface structure and adsorbate configurations, they do not capture the structural and chemical heterogeneity of practical catalysts. Accordingly, growing attention has shifted to polycrystalline[21–23] and nanoparticle (NP) structures[24–26] that more closely resemble industrial catalysts.

Platinum (Pt), with its near-optimal oxygen binding energy, lies near the apex of the ORR activity volcano plot and has been widely used as a benchmark ORR catalyst.[3] Yet, establishing structure–activity relationships on Pt remains challenging. Even in the absence of any dissolved molecular oxygen ($O_2$) in the electrolyte, the Pt surface undergoes many structural changes at ORR-relevant electrode potentials, including OH adsorption/desorption,[19,27–30] oxide formation/reduction,[19,27,28,31–33] surface roughening,[34] place exchange,[35,36] and/or Pt surface extraction.[35] The exact details of these surface evolutions, even for Pt single crystals, remain under debate to date.[29,31,37] In the presence of dissolved $O_2$, ORR produces additional oxygenated intermediates, further complicating the Pt surface structure.

*In situ* surface-sensitive spectroscopies, including infrared,[9,26,38] Raman,[39–44] and X-ray-based spectroscopy,[45,46] have been used to probe oxygenated species on Pt during ORR. In acidic electrolytes, the ORR intermediates observed on Pt single crystals vary with surface facet: adsorbed OOH ($OOH_{ad}$) on Pt(111),[39,44] adsorbed OH ($OH_{ad}$) on Pt(100) and Pt(110),[39] and both $OOH_{ad}$ and $OH_{ad}$ on Pt(211) and Pt(311).[44] For NPs with multiple Pt surface facets in acidic media, $OOH_{ad}$[26,40–43,46] and/or $OH_{ad}$[45,46] was observed in different measurements, which are mostly consistent with the single-crystal results. In near-neutral and weakly alkaline electrolytes (pH 5.5–11), existing studies of ORR at Pt single crystals and polycrystals have universally identified superoxide anion ($O_2^-$) as a key intermediate.[9,39,47]

Despite the tremendous existing efforts devoted to understanding the surface speciation of Pt under alkaline ORR conditions, a few key limitations and open questions remain. First, existing spectroscopy measurements only cover a pH up to 11, while a pH of 13 or higher is required under the operating conditions of AFCs and AAEMFCs. Second, studies of surface speciation of commercially relevant Pt NPs in alkaline ORR are still absent, to the best of our knowledge. Third, existing studies mainly showed spectroscopic results obtained over a single potential scan, missing the possible hysteresis and long-term evolution of surface structure. Fourth, existing works on alkaline ORR intermediates have not conducted control measurements of Pt in $O_2$-free electrolytes, making it difficult to isolate multiple contributing factors (inherent Pt surface process vs ORR process).

In this study, we use *in situ* electrochemical nanoparticle-enhanced Raman spectroscopy (EC-NERS) to study Pt surface evolution during alkaline ORR. The catalysts are NPs consisting of a gold (Au) core (for plasmonic enhancement) and a Pt shell (as ORR catalyst), which resemble the multifaceted commercial platinum/carbon (Pt/C) catalysts.[42,48–50] Previous studies used similar Au/Pt catalysts for EC-NERS during acidic ORR, and reported ORR activities comparable to those of commercial Pt/C.[41–43] Here we extend the study to alkaline ORR. We bridge the existing gap in knowledge by 1) performing EC-NERS in strong alkaline electrolyte with pH ~13; 2) measuring under different gas environments (Ar-saturation and $O_2$-saturation), through multiple potential scans, and in multiple electrolytes (KOH and LiOH); and 3) quantifying the potential- and history-dependent evolution of all the observed oxygenated species. In contrast to previous studies in weakly alkaline conditions for Pt single crystals and polycrystals,[9,39] we observed different ORR-produced surface adsorbates, including $OOH_{ad}$, $OH_{ad}$, and $O_{2,ad}$. Surprisingly, these species are hysteretic and highly long-lived on the scale of at least 1 hour, and are modestly correlated with the surface oxidation conditions of Pt and the electrolyte cations.

## RESULTS AND DISCUSSION

### Design and Characterization of the Experimental Platform

Figure 1a illustrates the working electrode configuration used for EC-NERS, in which Au/Pt NPs were deposited on a glassy carbon substrate. This configuration resembles the standard Pt/C catalysts widely used in commercial applications such as fuel cells, electrolyzers, and chemical manufacturing.[51–53] The UV-Visible spectra of Au NP solutions exhibited a localized surface plasmon resonance frequency at ~535 nm, corresponding to a particle size of ~60 nm (Figure S1), consistent with the scanning electron microscope (SEM) image showing spherical Au NPs with a diameter of 56.4±6.5 nm (Figure S2a). The spherical morphology was preserved after Pt shell coating (Figure S2b,c). Transmission electron microscope (TEM) images further exhibited characteristic lattice fringes of the Au core and Pt shell (Figure S2c,d). Using the L-shell emission of Au and Pt, scanning transmission electron microscopy energy dispersive X-ray spectroscopy (STEM-EDS) mapping and line profile were obtained, confirming that a ~2–3 nm thick Pt shell uniformly covers the Au core (Figure 1b,c).

To examine the capability of the Au/Pt NP assembly in enhancing Raman signal, we conducted three-dimensional finite-difference time-domain (3D-FDTD) simulation, and used $|E/E_0|^4$ to estimate the Raman enhancement factor (EF), where $E$ is the local electric field and $E_0$ is the incident electric field.[54] As shown in Figure 1d, we observed a plasmonic hotspot at the junction region between neighboring NPs, with an EF up to $4.7 \times 10^7$, consistent with previous reports.[55,56] These results reveal that the NERS signal primarily originate from the Pt surface rather than from Au or glassy carbon substrate. Guided by the 3D-FDTD results, we acquired NERS spectra from clustered NP regions to maximize the density of hot spots and thus the overall Raman signal from the Pt surface (Figure 1e). This strong Raman enhancement effect was further confirmed by comparing EC-NERS with the corresponding electrochemical Raman measurements of Pt/C (Figure S3).

Figure 1f shows cyclic voltammograms (CVs) of Au/Pt NPs in Ar-saturated 0.1 M LiOH and KOH aqueous solutions. In both electrolytes, the CVs exhibited characteristic Pt features similar to those observed for Pt/C (Figure S4a).[57,58] During the negative-going scan (NGS), the reduction of oxygenated adlayers ($OH_{ad}$ and $O_{ad}$) produced from Pt oxidation occurred at ~1.0–0.5 V, followed by hydrogen underpotential deposition at ~0.45–0.05 V. All potentials in this article are referenced to the reversible hydrogen electrode (RHE), unless otherwise noted. The positive-going scan (PGS) displayed hydrogen desorption at ~0.05–0.45 V and the formation of $OH_{ad}$ and $O_{ad}$ at ~0.65–1.0 V. In both solutions, the hydrogen adsorption/desorption peaks at ~0.25–0.28 V and ~0.36–0.40 V indicate the presence of Pt(110) and Pt(100) facets, respectively.[59–61] During the PGS, the $OH_{ad}$ and $O_{ad}$ formation feature showed cation-dependence, with a more pronounced anodic peak at ~0.82 V for $K^+$, whereas their reduction peak during the NGS was largely cation-independent. This agrees with the cation effects observed in the CVs of Pt/C (Figure S4a). Additionally, Au/Pt and Pt/C showed similar electrochemical behavior in acidic electrolyte, while Au NPs exhibited distinct redox peaks (Figure S5). These series of CV results confirmed the full, pinhole-free coverage of Pt on Au in the Au/Pt NPs, and the nearly equivalent electrochemical properties of our Au/Pt NPs vs the standard Pt/C catalysts.

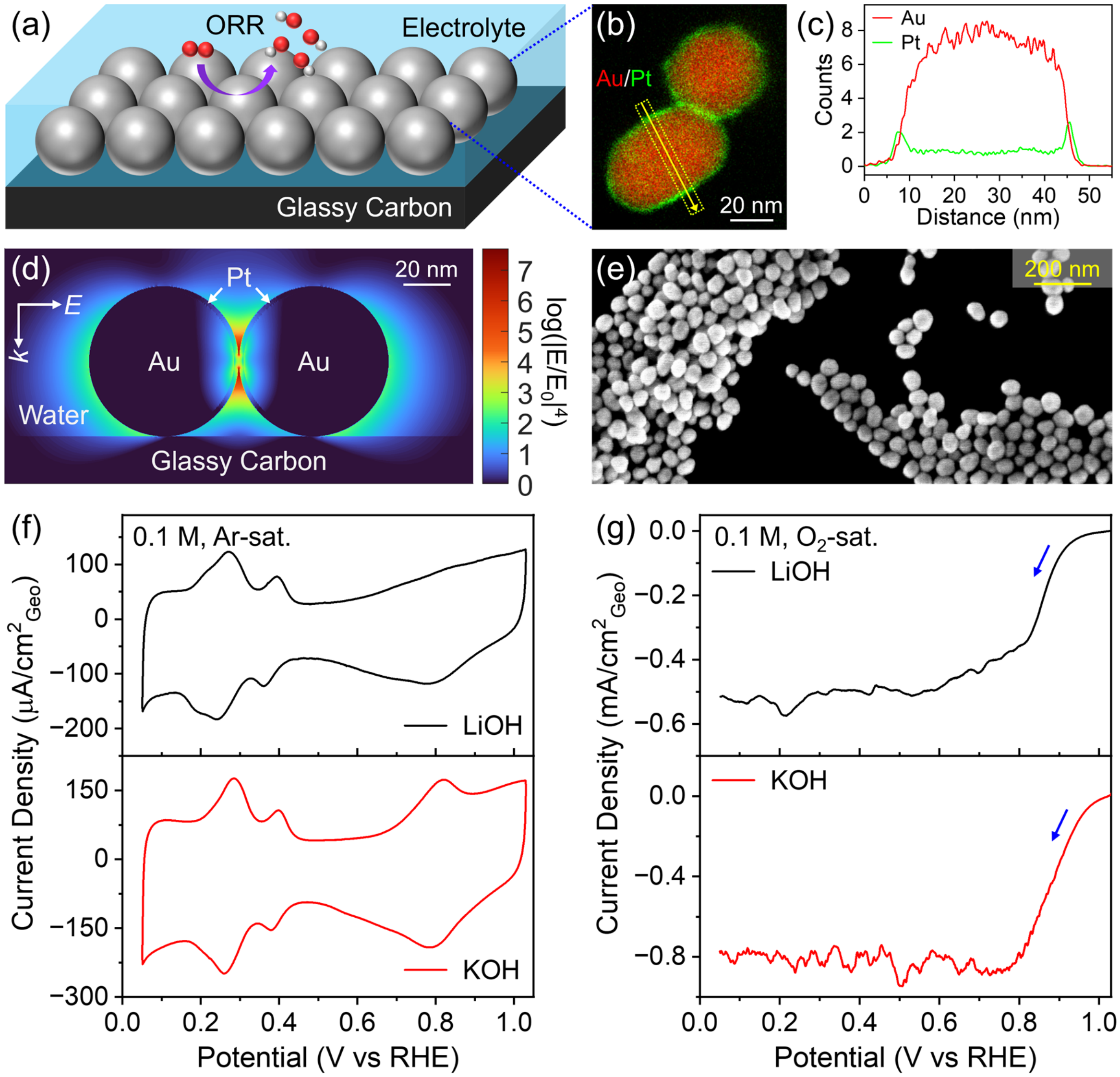

**Figure 1. Design and characterization of the *in situ* EC-NERS platform.** (a) Schematic of the working electrode in the EC-NERS measurements using Au/Pt NPs (O: red, H: light gray, Pt: gray). (b) STEM elemental map of Au/Pt NPs with Au in red and Pt in green, and (c) the corresponding line profile obtained from the region within the yellow dotted box, along the yellow arrow indicated in (b). (d) 3D-FDTD simulation showing the distribution of the Raman EF surrounding two Au/Pt NPs on a glassy carbon substrate in water. The polarization (*E*) and propagation (*k*) directions of the excitation laser are shown. (e) Scanning electron microscopy (SEM) image of Au/Pt NPs on a silicon substrate. (f) CVs of Au/Pt NPs in Ar-saturated 0.1 M LiOH and KOH aqueous solutions, measured at 100 mV/s. (g) ORR polarization curves of Au/Pt NPs in $O_2$-saturated 0.1 M LiOH and KOH aqueous solutions, measured under the NGS at 10 mV/s. Blue arrows indicate the scan direction. In (f,g), the geometric area of 0.20 $cm^2$ was used for current density conversion.

We further measured ORR polarization curves of Au/Pt NPs in $O_2$-saturated 0.1 M LiOH and KOH, obtained during the NGS (Figure 1g). The ORR onset potential ($E_{ORR}$) was defined as the potential at which the current density reaches −0.1 mA/cm$^2$.[62,63] $E_{ORR}$ was 0.89 V in LiOH and 0.95 V in KOH, indicating greater ORR activity in KOH. As the potential was reduced below $E_{ORR}$, the cathodic current density increased rapidly due to enhanced ORR kinetics with increasing overpotential, before reaching a plateau arising from mass transport limits. These observations are consistent with benchmark Pt/C behavior (Figure S4b).

The polarization curves in Figure 1g were collected without hydrodynamic control (i.e., electrode rotation) to reflect the static conditions of NERS measurements; therefore, the limiting current was lower than that obtained under rotating disk conditions (Figure S6a). Nevertheless, the more positive $E_{ORR}$ in KOH than in LiOH was preserved for both Au/Pt and Pt/C, regardless of the extent of hydrodynamic control (Figures 1g, S4b and S6).

Note that the current density measured in Figure 1f,g was sensitive to the surface coverage of the NPs, which fluctuates from sample to sample; as such, the exact current density values in each curve cannot be directly compared and used to infer electrolyte effects. A common practice to account for variations in surface coverage is normalization by the electrochemical surface area. While this approach is appropriate in kinetically controlled regimes, it cannot be applied in the mass transport-limited regime at high overpotentials, where the current is no longer proportional to the number of active sites.[64,65]

Through combined structural and electrochemical characterizations, together with electromagnetic simulation, we have demonstrated that the NERS platform enables direct probing of Pt surface chemistry while maintaining an industrially relevant catalyst structure and electrochemical behavior comparable to Pt/C.

**EC-NERS in Ar-Saturated Solution**

We obtained *in situ* EC-NERS spectra of Au/Pt in Ar-saturated 0.1 M KOH aqueous solution under the NGS (Figure 2a,b) and PGS (Figure 2c). Under all scans, a broad Raman band appeared at 1.0 V in the 320–700 cm$^{-1}$ range. This band likely consists of multiple Pt–O vibrational modes of platinum oxide ($PtO_x$), including stretching and libration vibrations, although the exact peak assignment and deconvolution are still under debate and beyond the scope of this work.[32,35,37,66–68] Previous studies revealed that amorphous $PtO_x$ exhibits a broad band, whereas crystalline $PtO_x$ displays sharp spikes.[31,32,68] In all scans, we identified amorphous $PtO_x$ as the dominant oxide structure. Note that the broad 320–700 cm$^{-1}$ band may also contain a small contribution from hydroxide species.[35] However, the extent of hydroxide contribution is debatable.[31,32,35,37,66,69] For simplicity, here we use "$PtO_x$" to denote all oxygenated species that may contribute to the 320–700 cm$^{-1}$ band; alternative assignments, including the existence of small amounts of hydroxides or other species, will not affect the subsequent conclusions we reached in this article.

We used the $PtO_x$ band area as a measure of oxide coverage on Pt and extracted its potential dependence (Figure 2d,e). In Ar-Scans 1 and 2 (NGS), the $PtO_x$ area decreased rapidly as the potential was decreased from 1.0 V to 0.4–0.5 V, before reaching a plateau at ~0.4 V and more negative potentials (Figure 2d,e). The potential range over which the $PtO_x$ area reduced is consistent with the CV data, where the $PtO_x$ reduction feature persisted to ~0.5 V under the NGS (Figure 1f). Ar-Scan 3—the subsequent PGS right after Ar-Scan 2—exhibited an extended reduced-state plateau over 0.0–0.7 V, followed by a small increase in the $PtO_x$ band area between 0.7 and 1.0 V (Figure 2e). This is also consistent with the corresponding CV result, where the formation of oxygenated adlayers started at E>~0.65 V under the PGS (Figure 1f). Comparing Ar-Scans 2 and 3, we observed comparable $PtO_x$ areas in the 0.0–0.4 V plateau region, whereas Scan 3 exhibited a lower $PtO_x$ area over the 0.5–1.0 V range, indicating hysteresis in the potential-induced evolution of oxide coverage on Pt (Figure 2e).

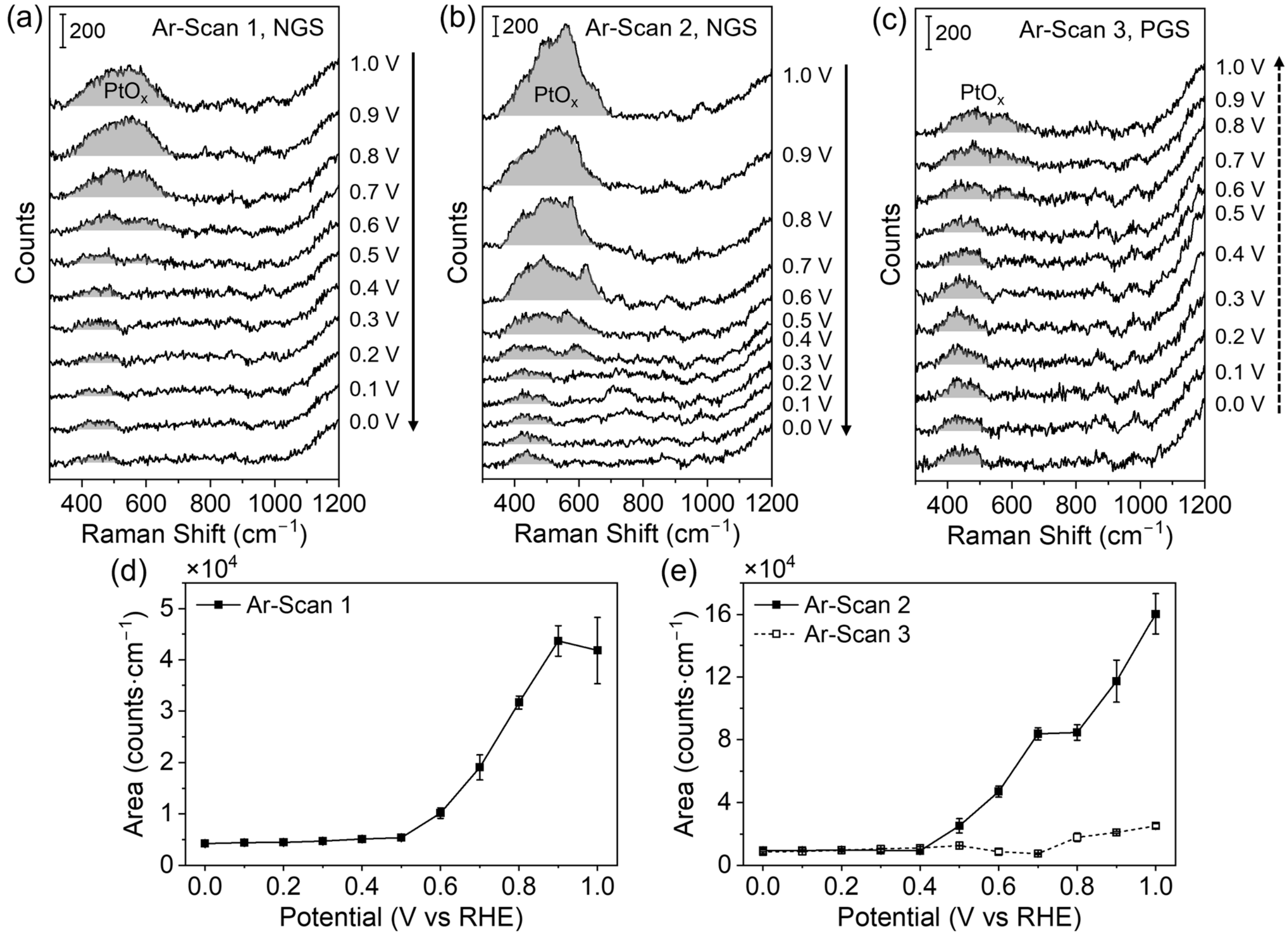

**Figure 2. Potential-dependent evolution of oxide on Pt nanocatalysts in Ar-saturated 0.1 M KOH.** *In situ* EC-NERS spectra of Au/Pt NPs obtained in Ar-saturated 0.1 M KOH aqueous solution during (a,b) NGS and (c) PGS right after (b). The gray-shaded regions mark the $PtO_x$ band area. (d,e) Potential-dependent evolution of the $PtO_x$ band area extracted from (a) and (b,c), respectively. In (e), solid and dashed lines correspond to results extracted from (b) and (c), respectively. Statistics were derived from technical replicates. The broad spectral uplift in the 1050–1200 $cm^{-1}$ region arises from the disorder-induced (D) band of the glassy carbon substrate.[70,71]

Note that the $PtO_x$ area was above zero in the reduced-state plateaus in all the three scans, although this area was close to the noise baseline. In this regime, small amounts of $PtO_x$ or $OH_{ad}$ could exist on polycrystalline Pt and at defective sites such as step edges, kinks, vacancies, and dislocation sites.[72–74] Indeed, $OH_{ad}$ was found to be absent on atomically flat Pt(111) terraces but present on non-Pt(111) facets and step edges at potentials below 0.4 V both in acid and alkaline media.[29,75,76] The presence of $PtO_x$ species on Au/Pt NP surfaces is therefore possible due to their multi-facet character and the significantly larger amounts of defect sites of the more practical NP catalysts. Interestingly, our EC-NERS measurements in acid solution revealed the presence of $OH_{ad}$ at <0.4 V (Figure S3b), consistent with the previous report on step-edge systems.[29] Overall, due to the stronger heterogeneity of the NP catalysts compared to their single-crystal counterparts, we expect significant differences in the nature and evolution of surface species under the same electrochemical conditions.

## EC-NERS in $O_2$-Saturated Solution

To investigate the Pt surface speciation under ORR conditions, we proceeded to conduct EC-NERS measurements in $O_2$-saturated 0.1 M KOH aqueous solution (Figure 3). In all the potential scans, we observed a $PtO_x$ band in the 320–700 $cm^{-1}$ range. The $PtO_x$ band area decreased as the electrode potential was reduced from 1.0 to 0.6–0.7 V, before reaching a reduced-state plateau at more negative potentials. The $PtO_x$ band position and potential-dependent evolution were similar to the results obtained in Ar saturation.

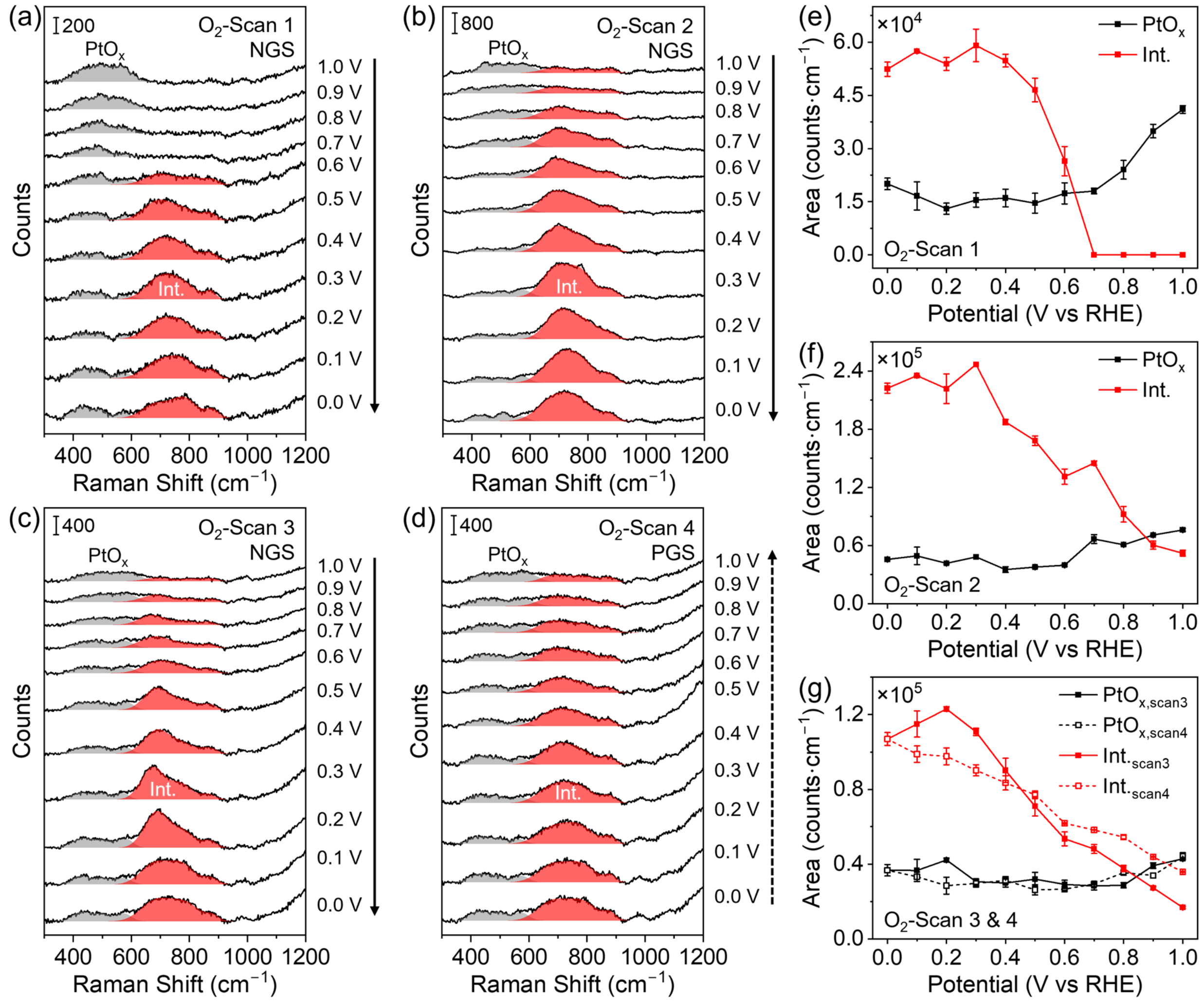

**Figure 3. Potential-dependent evolution of Pt surface species in $O_2$-saturated 0.1 M KOH.** *In situ* EC-NERS spectra of Au/Pt NPs obtained in $O_2$-saturated 0.1 M KOH aqueous solution during (a–c) NGS and (d) the subsequent PGS right after (c). The gray- and red-shaded regions denote

the band areas of $PtO_x$ and ORR intermediates, respectively. An elevation in the 1050–1200 $cm^{-1}$ region originates from the glassy carbon D-band.[70,71] (e–g) Potential-dependent evolution of $PtO_x$ (black) and ORR intermediate (red) band areas, extracted from (a), (b), and (c,d), respectively. In (g), solid and dashed lines correspond to results extracted from (c) and (d), respectively. Statistics were derived from technical replicates.

Compared to the EC-NERS results in Ar-saturated electrolyte, we observed an additional Raman band between ~550 and 950 $cm^{-1}$ in $O_2$-saturated conditions. Unlike the $PtO_x$ band, which tended to decay before plateauing at more negative potentials, the 550–950 $cm^{-1}$ band generally increased before plateauing at lower potentials (Figure 3e–g). Considering this potential-dependent trend, along with the fact that the 550–950 $cm^{-1}$ band only emerged in the presence of $O_2$, we attribute this band to intermediates produced by ORR. This band area was used as a measure of ORR intermediate coverage.

In $O_2$-Scan 1, the ORR intermediate band was initially absent, and started to emerge after the potential reached 0.6 V (Figure 3a,e). This onset potential, while below that of the global ORR onset (Figure 1g), matches with the potential at which the $PtO_x$ band was reduced to the minimum in the same scan at the local sample region measured by NERS (Figure 3e). This is likely due to the inactivity of $PtO_x$ sites for ORR, as reported before;[23,33] as a result, ORR intermediates reached NERS-detectable coverage only when the $PtO_x$ coverage was sufficiently low. As the potential reached ~0.4 V, the intermediate band area reached a plateau and no longer increased at more negative potentials, indicating a maximum coverage of ORR intermediates. After $O_2$-Scan 1, the Pt / $O_2$-saturated 0.1 M KOH system was left at open circuit potential (OCP, ~0.96 V) for ~81 minutes before EC-NERS was measured again. As shown in Figure S7a, a small intermediate band was still present after the termination of reducing potentials for ~81 minutes, revealing the long-lived nature of these surface species.

At the beginning of $O_2$-Scans 2 and 3, small amounts of ORR intermediates were observed (Figure 3b,c,f,g), due to the prior history of ORR. In contrast to $O_2$-Scan 1, where the intermediate band did not increase until 0.6 V, $O_2$-Scans 2 and 3 both featured an immediate increase in the intermediate band area as soon as the potential was reduced below 1.0 V. This is likely again due to the competitive accumulation effects of $PtO_x$ vs ORR intermediates. More specifically, the initial presence of ORR intermediates on a fraction of the Pt surface area ensures the absence of $PtO_x$ in these areas; as such, these intermediate-covered areas are more active for ORR and enable more positive onset potentials for the increase in intermediate coverage. As the potential became more negative, both $O_2$-Scans 2 and 3 showed saturation of the intermediate band area at 0.3 V and lower, likely due to full coverage of the available surface sites, similar to $O_2$-Scan 1.

Right after $O_2$-Scan 3, we conducted $O_2$-Scan 4 (PGS, Figure 3d,g). As the potential increased from 0.0 V, we observed a gradual decrease in the band area for ORR intermediates. As the potential reached 1.0 V, the intermediate band was still finite and higher than the original 1.0 V result in $O_2$-Scan 3. These results reveal the hysteretic nature of the observed ORR intermediates. In a separate scan over the same electrode area, after the adsorption of ORR intermediates, the Pt surface was intentionally oxidized by holding the potential at 1.1 V for ~22 minutes, followed by 1.0 V for ~17 minutes; afterwards, the intermediate band was still present (Figure S7a), further confirming their hysteretic nature.

To further verify that the potential-dependent, hysteretic evolution of the intermediate band was due to ORR on Pt, instead of other side effects, we conducted a series of control experiments. First, CV measurements of Au/Pt NPs in Ar-saturated solution before and after ~5 hours of ORR (0.4 V, $O_2$-purged, 0.1 M KOH) revealed nearly the same characteristic Pt redox peaks and negligible Au redox peaks, confirming the stability of the Pt shell and absence of pinholes / exposed Au during ORR (Figure S8). Second, control EC-NERS measurements using bare Au NPs under ORR conditions revealed distinct peak features compared to those of Au/Pt NPs, further validating the full coverage of Pt shell of the Au/Pt NPs during EC-NERS (Figure S9). Third, Raman peaks of the glassy carbon substrate before and after ~3.5 hours of EC-NERS under ORR conditions remained largely identical, confirming the lack of carbon oxidation or corrosion effects (Figure S10). These three control experiments together confirmed that the EC-NERS results using Au/Pt NPs during ORR were not due to the Au core exposure or glassy carbon degradation.

**Intermediate Deconvolution and Reaction Pathway**

The broad distribution of the intermediate band (~550–950 $cm^{-1}$) in $O_2$-Scans 1–4 indicated the coexistence of multiple vibrational modes. To elucidate the nature of these modes, we deconvoluted this band via Gaussian peak fitting, and identified three peak components within each band (Figure 4a–d; details in Supporting Information, Materials and Methods 3.7). The peaks at ~686 and ~875 $cm^{-1}$ were attributed to the O–O stretching modes ($\nu$(O–O)) of $OOH_{ad}$[39,41,77] and $O_{2,ad}$,[77,78] respectively, while the peak at ~778 $cm^{-1}$ was assigned to the Pt–O–H bending mode ($\delta$(Pt–O–H)) of $OH_{ad}$.[79] A previous study confirmed these assignments by the $D_2O$ isotopic shift.[77] In addition, $O_{2,ad}$ is likely in a bridge-sorbed configuration,[1,80] as previous vibrational frequency calculations showed that $\nu$(O–O) of $O_{2,ad}$ appears above 1200 $cm^{-1}$ for end-on adsorption on Pt[39,78] and below 1000 $cm^{-1}$ for a bridge-sorbed configuration.[77,78] The exact adsorption configuration of $OOH_{ad}$—end-on or bridge—remains under debate,[39,40,42,77] although the conclusions of this work will not depend on such precise assignment.

Figure 4e displays the potential-dependent areas of the deconvoluted components ($OOH_{ad}$, $OH_{ad}$, $O_{2,ad}$) for each scan. In all scans, both $OOH_{ad}$ and $OH_{ad}$ exhibited larger areas in the low potential range (0.0–0.5 V) compared to those above 0.5 V. This observation appears to conflict with existing microkinetic models, which predict near-zero surface coverage of ORR intermediates at high overpotentials ($E<0.7$ V), where reaction kinetics is more facile than mass transport.[1,81] Although these models successfully reproduce experimental current density–potential characteristics,[1,82] the discrepancy with our EC-NERS results indicates that the models may not fully capture all the oxygenated species produced during ORR on NP catalysts. Furthermore, at 1.0 V, where ORR is absent, the $OOH_{ad}$ and $OH_{ad}$ areas remained finite in $O_2$-Scans 2–4 due to the prior ORR history, with larger areas observed in $O_2$-Scan 4 compared to those in $O_2$-Scan 3. Additionally, the area of $O_{2,ad}$ also remained above zero throughout all the potentials in $O_2$-Scans 2–4, although the area was close to the noise baseline at potentials close to 1.0 V. These results suggest the accumulation of kinetically inactive intermediates during ORR, which can account for the apparent inconsistency with simplified microkinetic models that do not consider such species.

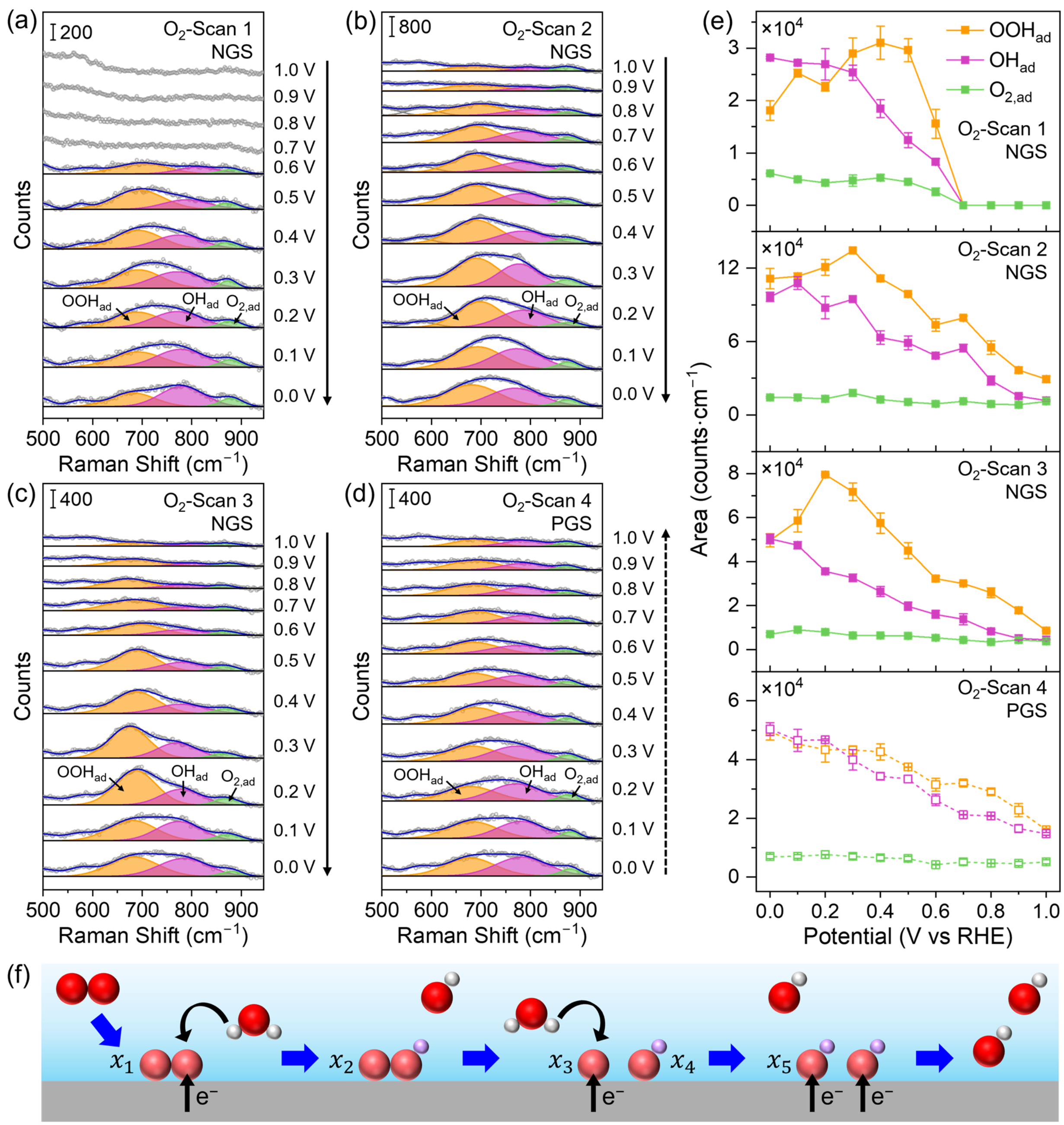

**Figure 4. Deconvolution of ORR intermediate species.** Deconvoluted EC-NERS spectra of Au/Pt NPs in $O_2$-saturated 0.1 M KOH aqueous solution during (a–c) NGS and (d) the subsequent PGS right after (c). Gray scattered points are the raw data. The orange, magenta-purple, and green shaded areas correspond to the $\nu$(O–O) mode of $OOH_{ad}$, $\delta$(Pt–O–H) mode of $OH_{ad}$, and $\nu$(O–O) mode of $O_{2,ad}$, respectively. The black line represents the baseline, while the navy line corresponds to the sum of all fitting components, which consist of mainly ORR-induced modes and a small contribution from $PtO_x$. (e) Potential-dependent evolution of the areas of $OOH_{ad}$ (orange), $OH_{ad}$ (magenta-purple), and $O_{2,ad}$ (green) components during $O_2$-Scans 1–4 (top to bottom) extracted from (a–d). (f) Schematic of proposed ORR steps on Pt nanocatalysts in alkaline media (O atom of non-adsorbed species: red, H atom of non-adsorbed species: light gray, O atom of adsorbed species: light red, H atom of adsorbed species: light purple).

Building upon the associative mechanism for alkaline ORR proposed in previous studies,[1,19] and considering the distinction between active and inactive intermediates, we propose the following ORR pathway of Pt nanocatalysts in alkaline media (Figure 4f):

$$O_2 \rightarrow x_1 O_{2,ad,active} + (1 - x_1) O_{2,ad,inactive} \quad (1)$$

$$O_{2,ad,active} + H_2O + e^- \rightarrow x_2 OOH_{ad,active} + (1 - x_2) OOH_{ad,inactive} + OH^- \quad (2)$$

$$OOH_{ad,active} \rightarrow x_3 O_{ad,active} + (1 - x_3) O_{ad,inactive} + x_4 OH_{ad,active} + (1 - x_4) OH_{ad,inactive} \quad (3)$$

$$O_{ad,active} + H_2O + e^- \rightarrow x_5 OH_{ad,active} + (1 - x_5) OH_{ad,inactive} + OH^- \quad (4)$$

$$2 \times [OH_{ad,active} + e^- \rightarrow OH^-] \quad (5)$$

where $x_i$ ($i$ = 1–5) is the fractional factor of the corresponding active species produced at specific steps. The dissociative pathway is not considered here, as it does not generate $OOH_{ad}$.[1,83] Additionally, our EC-NERS results revealed that the water peak was potential-independent (Figure S11); as such, we considered water to be non-adsorbed in the reaction process.

In addition to the reaction steps described above, another ORR pathway is possible, where eq 3 involves electron transfer and the production of $OH^-$, as summarized in Note S1. While we cannot identify the exact molecular pathways of the ORR process, the possible small variations will not affect the main conclusions we reach in this work.

In $O_2$-Scans 1–3 (NGS), the peak area of $OOH_{ad}$ first increased before decreasing at sufficiently negative potentials; in contrast, $OH_{ad}$ either first increased before plateauing ($O_2$-Scans 1 and 2) or continuously increased throughout the whole potential range ($O_2$-Scan 3) (Figure 4e). Considering that kinetically active intermediates are expected to decay rapidly in the mass-transfer limited regime, these results indicate that $OH_{ad,inactive}$ likely dominates over $OH_{ad,active}$ in eqs 3 and 4. As to $OOH_{ad}$, the slight decay at negative potentials reveals that both the $OOH_{ad,active}$ and $OOH_{ad,inactive}$ are produced in eq 2. In $O_2$-Scan 4 (PGS), as the potential increased from 0.0 V, both $OOH_{ad}$ and $OH_{ad}$ areas decayed nearly continuously. This decay suggests that some of the inactive intermediates converted into active intermediates and proceeded through the following ORR steps. As to the $O_{2,ad}$ species in $O_2$-Scans 1–4, their weak signal prevents a precise quantification of their potential- and history-dependent evolution.

Our observed kinetically inactive intermediates likely originate from the structural heterogeneity of the nanocatalysts. Previous studies reported more negative binding energies and higher maximum coverage of oxygenated species at Pt step sites compared to terrace sites,[73,84] suggesting that stronger binding at defect-rich surfaces of nanocatalysts may facilitate the accumulation of intermediates observed in this work. Consistent with this, a longer residence time of intermediates was predicted on Pt surfaces with higher $O_2$–Pt coordination numbers.[15] Similar effects have been reported on copper surfaces during CO or $CO_2$ reduction: defective sites bind intermediates more strongly, while nanocavities enrich intermediates, increasing their surface coverage and residence time.[85–87] The morphology-dependence of intermediate coverage is likely due to our recently reported solid–liquid superposition effect:[88] the local corrugation of catalyst surface, particularly at concave sites, enables higher local density of intermediates due to the linear superposition of the intermediate–catalyst interaction potential.

Note that ORR activity on high-index plane Pt with stepped sites is known to be greater than on Pt(111) in acidic conditions, whereas the opposite trend has been observed in alkaline conditions.[18,83,89] Previous studies of Pt-covered NP catalysts during acidic ORR revealed a decay of the intermediate peak area at negative potentials and absence of intermediates as the potential was close to 0.0 V.[40–42] This implies that the accumulation of kinetically inactive intermediates at Pt defect sites is likely more favorable in alkaline conditions than in acidic media.

Our identified intermediates differ from those reported in prior studies on both single-crystal and polycrystalline Pt under weakly alkaline media of pH 10–11, where $O_2^{-}{}_{,ad}$ species were observed in the 1000–1200 $cm^{-1}$ range (Note S2).[9,39] This difference may be due to the more defective surface structure of the NP catalysts in our measurements. Future studies to systematically examine the effects of pH and Pt surface structure on ORR intermediates will be highly valuable, but beyond the scope of this work.

**Electrolyte Effect**

Over the past decade, the impact of electrolyte species on electrocatalytic activities has been extensively studied.[8,19,90–92] For Pt-catalyzed ORR, while alkali cations are known to modulate the catalytic activity, direct insight into the cation-dependence of Pt surface speciation during ORR remains limited.[8,19,27,28,35] To investigate electrolyte effects, we acquired NERS spectra in Ar- and $O_2$-saturated 0.1 M LiOH aqueous solution (Figures S12 and S13), and compared the results with those in 0.1 M KOH aqueous solution (Figures 2–4). In both Ar- and $O_2$-saturated 0.1 M LiOH, we observed the $PtO_x$ band in the 320–700 $cm^{-1}$ range. Under $O_2$ saturation, intermediate peaks, identified as the $\nu$(O–O) mode of $OOH_{ad}$, the $\delta$(Pt–O–H) mode of $OH_{ad}$, and the $\nu$(O–O) mode of $O_{2,ad}$, occurred in the ~550–950 $cm^{-1}$ range. These observations are similar to those in 0.1 M KOH (Figures 2–4).

In Ar-Scans 4 and 5 (LiOH, NGS), the $PtO_x$ coverage decreased and reached a reduced-state plateau as the potential became more negative (Figure 5a,b), similar to the KOH results (Figure 2). Ar-Scan 6—the PGS following Ar-Scan 5—exhibited a nearly constant $PtO_x$ area as the potential increased, indicating hysteretic evolution of $PtO_x$ (Figure 5b,c). This PGS trend is different from that in KOH (Ar-Scan 3), where the $PtO_x$ area increased from 0.7 V to 1.0 V (Figure 2e). The comparison of EC-NERS results in LiOH vs KOH is consistent with the CV comparisons of Pt in LiOH vs KOH solutions under Ar-saturation: for both Au/Pt and Pt/C, the $PtO_x$ reduction peak under the NGS were nearly the same in LiOH and KOH; in contrast, a sharp anodic spike at ~0.82–0.85 V occurred in KOH during the PGS, which was absent in LiOH (Figures 1f and S4a).

For $O_2$-Scans 5–7 (LiOH, NGS), the $PtO_x$ area exhibited an overall weak decay before plateauing at more negative potentials (Figure 5d–f), similar to that in KOH (Figure 3e–g). As to the total area of ORR intermediates, the initial potential-dependent trends at 1.0–0.3 V for $O_2$-Scans 5–7 (Figure 5d–f) were mostly the same as those in $O_2$-Scans 1–3 (Figure 3e–g), indicating a similar competitive accumulation effect of $PtO_x$ vs ORR intermediates as well as similar hysteresis of the intermediates, independent of the electrolyte cation. However, at potentials lower than ~0.3 V, the total intermediate areas in LiOH exhibited sharp decays (Figure 5d–f), in contrast to the nearly constant areas in KOH (Figure 3e–g). As discussed before, kinetically active intermediates are expected to decay at more negative potentials in the mass-transfer limited regime. As such, ORR in LiOH solution likely produced a larger fraction of active intermediates compared to ORR in KOH.

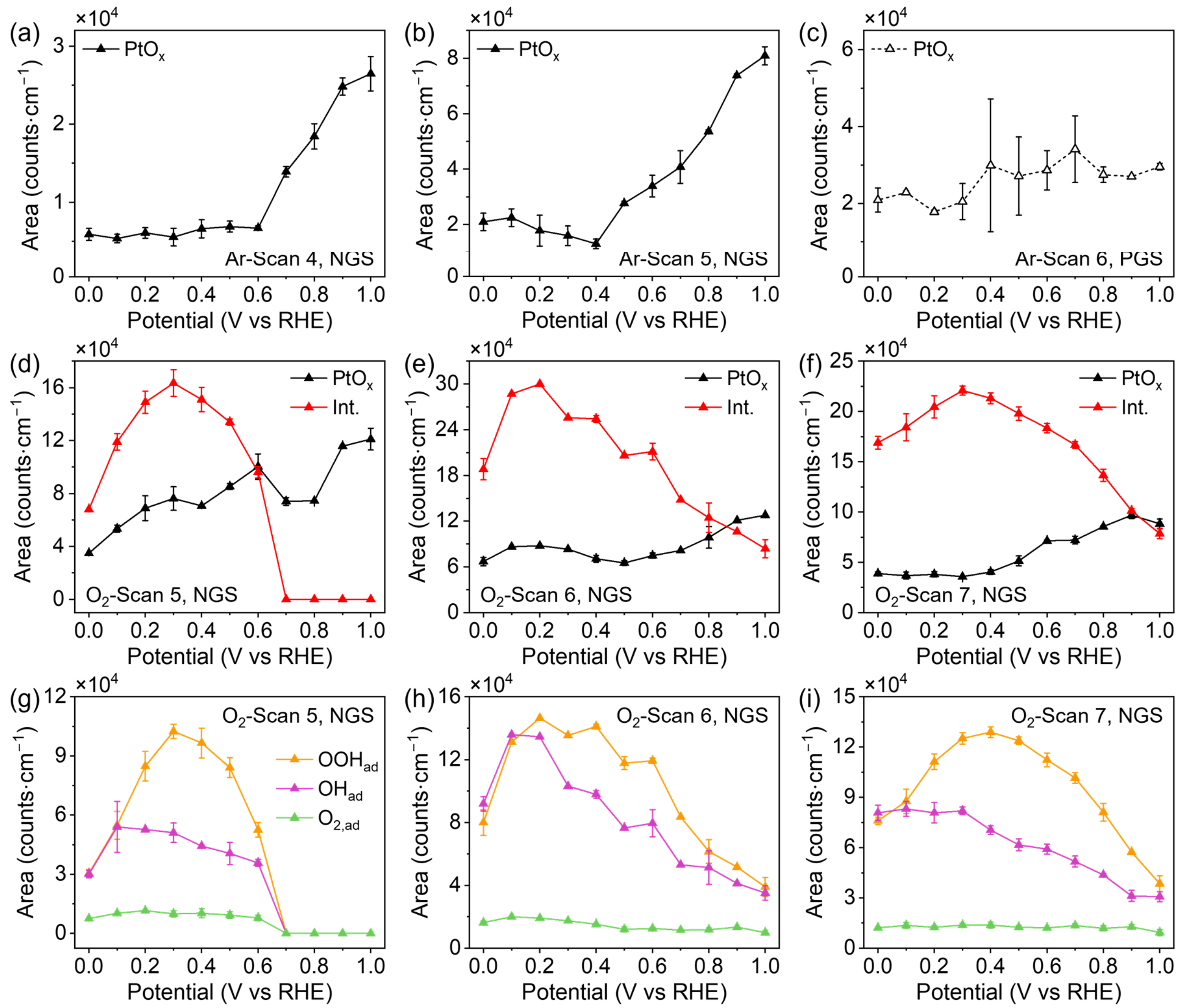

**Figure 5. Potential-dependent evolution of oxide and ORR intermediates on Pt in Ar- and $O_2$-saturated 0.1 M LiOH.** Potential-dependence of the $PtO_x$ band area in Ar-saturated 0.1 M LiOH aqueous solution during (a,b) NGS and (c) the subsequent PGS of (b). (d–f) Potential-dependence of $PtO_x$ (black) and ORR intermediate (red) band areas in $O_2$-saturated 0.1 M LiOH under NGS. (g–i) Potential-dependent areas of the $\nu$(O–O) mode of $OOH_{ad}$ (orange), the $\delta$(Pt–O–H) mode of $OH_{ad}$ (magenta-purple), and the $\nu$(O–O) mode of $O_{2,ad}$ (green), acquired during NGS in $O_2$-saturated 0.1 M LiOH. (a–c) were extracted from Figure S12, and (d–i) were extracted from Figure S13. Statistics were obtained from technical replicates.

The deconvoluted ORR intermediate areas in LiOH (Figure 5g–i) also exhibited nearly the same potential-dependent trends as those in KOH (Figure 4e) in the 1.0–0.3 V range. In LiOH solution, below 0.3 V, we observed a sharp decay in the $OOH_{ad}$ peak area, while the $OH_{ad}$ and $O_{2,ad}$ areas either remained mostly constant or slightly decayed. This indicates that a significant amount of $OOH_{ad}$ produced from reaction eq 2 was likely kinetically active, and was consumed quickly through eq 3 (greater $x_2$ in LiOH than in KOH). The observed fraction of kinetically active species might seem counterintuitive, given the positive shift of ~0.06 V in $E_{ORR}$ for KOH vs LiOH electrolytes in the polarization curves (Figure 1g). This could be explained by the blockage of defect sites by alkali metal cations, consistent with a previous study.[89] $Li^+$ is likely more effective than $K^+$ at blocking defect sites, possibly due to its smaller size.[93] These defect sites may have dual functions: 1) at potentials close to $E_{ORR}$, they may be active for ORR; and 2) at more negative potentials, due to weaker repulsion or stronger attraction with the cations, ORR intermediates at such sites may become strongly trapped and thus inactive. Therefore, more effective blockage of defect sites by $Li^+$ can both reduce the ORR activity at potentials close to $E_{ORR}$ and lower the fraction of kinetically inactive species at more negative potentials.

We further observed the retention of the ORR intermediate band for ~112 minutes after the termination of reducing potentials in LiOH (Figure S7b). Additionally, we conducted NERS measurements in $O_2$-saturated 0.1 M $NaClO_4$ + 0.1 M NaOH aqueous solution (Figure S9b,c), and observed similar $PtO_x$ and ORR intermediate bands and their potential dependence, compared to those in LiOH (Figures 5d–i and S9e,f). These results reveal that, while cations modulate the site-blocking effect, the overall existence and hysteresis of Pt surface species during alkaline ORR are intrinsic properties of the NP catalysts and occur universally, regardless of the cation species.

**Oxygenated Species Coverage on Pt Nanocatalysts and Mechanistic Implications**

The overall observations of this work are summarized in Figure 6. Our findings reveal hysteretic behavior and surface trapping effects of oxygenated species, including surface oxides and ORR intermediates, on Pt nanocatalysts in alkaline media. In Ar-saturated electrolytes, surface oxide reduction occurs predominantly in the high potential regime of 0.5–1.0 V, with a small residual oxide coverage persisting below 0.5 V, likely at defective sites (Figures 2 and 6a). In $O_2$-saturated conditions, a competitive relationship between $PtO_x$ and ORR intermediates emerges in the high potential regime of 0.6–0.9 V, while surface trapping of oxygenated species occurs at lower potentials (Figures 3 and 6b). The surface trapping effect leads to the coexistence of kinetically active and inactive ORR intermediates, with the inactive species exhibiting longer surface residence time (Figures 4 and 6b). The fractions of active and inactive intermediates are modulated by electrolyte cations, likely through cation blocking of defective surface sites that trap the inactive species (Figures 4, 5 and 6c).

The identification of kinetically inactive intermediates provides important guidance for interpreting vibrational spectroscopic signals under catalytic conditions. In particular, the conventional practice of directly correlating spectroscopic features with kinetic current responses may lead to misinterpretation if inactive species are present.[94] Furthermore, in realistic catalytic systems, surface oxides and kinetically inactive reaction-induced adsorbates may persist over extended periods of time before being further reduced or converted. Although distinguishing between kinetically active and inactive species in detail is beyond the scope of this work, further studies may gain deeper mechanistic insight by employing *in situ* spectroscopy measurements under alternating current or pulsed electrode potential modulation. Our observations highlight the significance of accounting for differences in kinetic activity of surface species when evaluating catalytic reaction mechanisms.

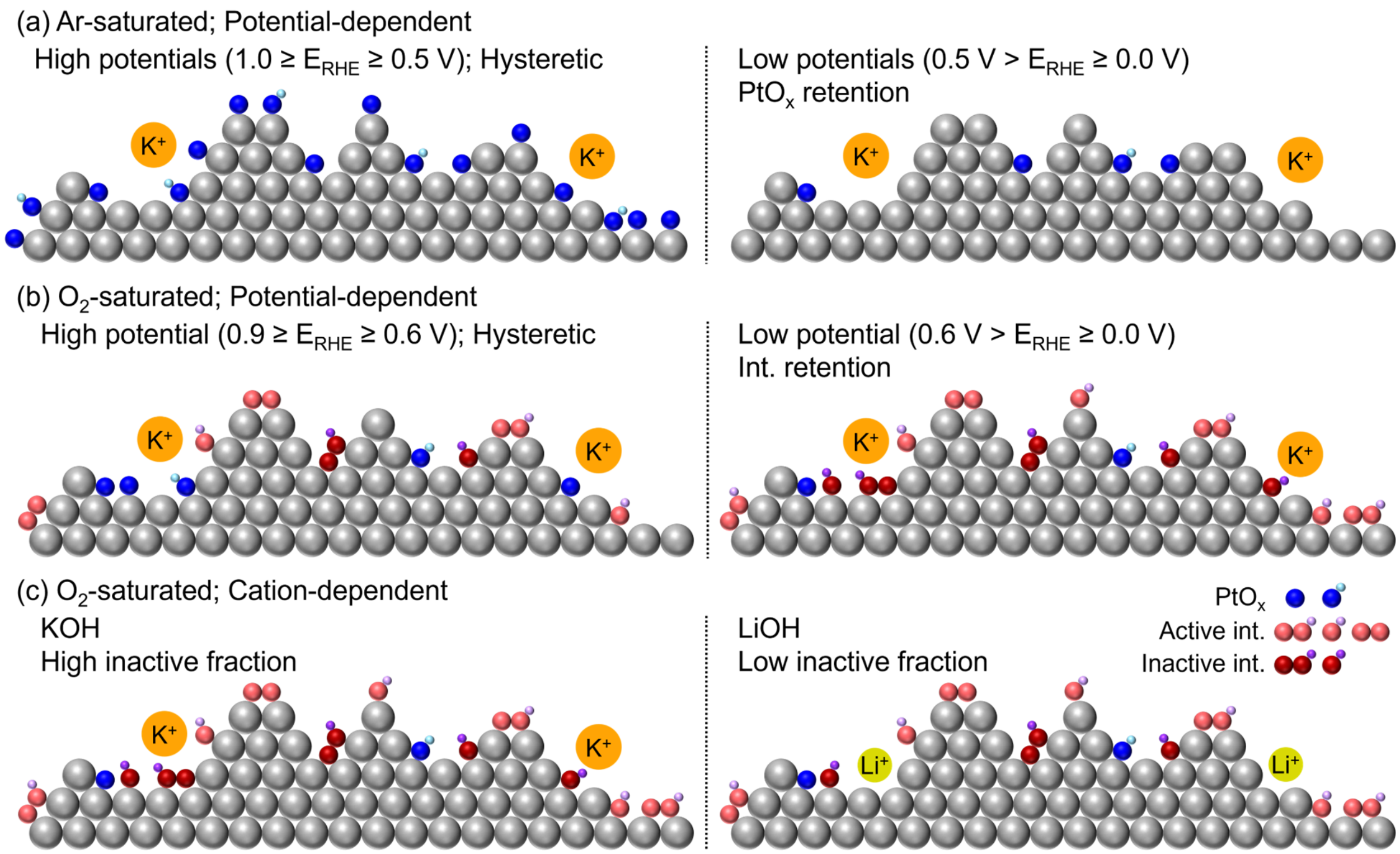

**Figure 6. Schematic illustration of surface speciation on Pt nanocatalysts in alkaline solutions.** (a) Potential-dependent oxygenated species coverage in Ar-saturated electrolytes with $PtO_x$ retention at defective sites. (b) Potential-dependent oxygenated species coverage in $O_2$-saturated electrolytes with the retention of $PtO_x$ and ORR intermediates at defective sites. (c) Cation-dependent coverage of oxygenated species in $O_2$-saturated electrolytes, with cation identity controlling the fraction of inactive intermediates trapped at defectives sites.

## CONCLUSIONS

In conclusion, we have resolved the surface speciation of Pt nanocatalysts by *in situ* electrochemical NERS in strongly alkaline electrolytes (pH ~13) under systematically controlled conditions, including electrolyte saturation gas (Ar vs $O_2$), multiple potential scans, and electrolyte species (mainly $K^+$ vs $Li^+$). We observed hysteretic, potential-dependent evolution of oxygenated species on Pt in the absence and presence of ORR. We identified the coexistence of kinetically active and inactive intermediates, with the inactive species exhibiting long-lived character and modest cation dependence. Surface species retention in both Ar- and $O_2$-saturated electrolytes and cation effect on intermediate coverage are all likely facilitated by the defect-rich structure of Pt nanocatalysts. These findings underscore the intricate nature of surface speciation on realistic nanocatalysts under reactive conditions, highlighting the importance of considering such intricacies when evaluating practical electrocatalytic systems.

## ASSOCIATED CONTENT

### Supporting Information

The Supporting Information is available free of charge online.

Materials and methods, alternative proposed pathway of ORR on Pt nanocatalysts in alkaline media, comparisons with previously reported ORR intermediates on Pt in weakly alkaline media, extinction spectra of Au NPs, electron microscopy images of Au and Au/Pt NPs, *in situ* EC-Raman spectra in acidic medium, electrochemical characterizations of Pt/C in alkaline media, CVs of Pt and Au nanocatalysts in acidic media, ORR polarization curves under hydrodynamic control, hysteretic adsorption of ORR intermediates on Pt nanocatalysts, electrochemical stability of Au/Pt NPs, *in situ* EC-NERS spectra of Au and Pt nanocatalysts in an $O_2$-saturated $Na^+$-based electrolyte, chemical stability after the NERS measurement, potential-independent behavior of the water peak, potential-dependent evolution of oxide on Pt in Ar-saturated 0.1 M LiOH, potential-dependent evolution of Pt surface species in $O_2$-saturated 0.1 M LiOH and deconvolution of ORR intermediate species (PDF)

**Notes**

The authors declare no competing financial interest.

**ACKNOWLEDGMENTS**

We acknowledge the support from the Beckman Young Investigator Award provided by the Arnold and Mabel Beckman Foundation and the Sloan Research Fellowship from the Alfred P. Sloan Foundation. J.K. acknowledges support from the TechnipFMC Educational Fund Fellowship and a PPG-MRL Graduate Research Assistantship program. The experiments were performed in part in the Beckman Institute for Advanced Science and Technology and in the Materials Research Laboratory at the University of Illinois Urbana-Champaign. We acknowledge Dr. Changqiang Chen for his help with STEM measurements.

**REFERENCES**

(1) Liu, S.; White, M. G.; Liu, P. Mechanism of Oxygen Reduction Reaction on Pt(111) in Alkaline Solution: Importance of Chemisorbed Water on Surface. *J. Phys. Chem. C* **2016**, *120* (28), 15288–15298. https://doi.org/10.1021/acs.jpcc.6b05126.

(2) Briega-Martos, V.; Herrero, E.; Feliu, J. M. Effect of pH and Water Structure on the Oxygen Reduction Reaction on Platinum Electrodes. *Electrochim. Acta* **2017**, *241*, 497–509. https://doi.org/10.1016/j.electacta.2017.04.162.

(3) Seh, Z. W.; Kibsgaard, J.; Dickens, C. F.; Chorkendorff, I.; Nørskov, J. K.; Jaramillo, T. F. Combining Theory and Experiment in Electrocatalysis: Insights into Materials Design. *Science* **2017**, *355* (6321), eaad4998. https://doi.org/10.1126/science.aad4998.

(4) Do, V.; Lee, J. Transforming Adsorbate Surface Dynamics in Aqueous Electrocatalysis: Pathways to Unconstrained Performance. *Adv. Mater.* **2025**, *37* (10), 2417516. https://doi.org/10.1002/adma.202417516.

(5) Chen, Z.; Lin, Z.; Zhu, X.; Li, Y.; Wang, Y. Adsorbed Oxygen Dynamics at Forced Convection Interface in the Oxygen Evolution Reaction. *Nat. Commun.* **2025**, *16* (1), 7949. https://doi.org/10.1038/s41467-025-63181-z.
(6) Ni, W.; Liang, Y.; Cao, Y.; Chen, Z.; Miao, R. K.; Peng, B.; Liu, Z.; Liu, Y.; Ze, H.; Wang, X.; Kim, D.; Park, S.; Yu, J.; Papangelakis, P.; Boureau, V.; Imran, M.; Wang, Q.; Ou, P.; Li, X.-Y.; Xie, K.; Dorakhan, R.; Shirzadi, E.; Schatz, G. C.; Sinton, D.; Ge, J.; Zeng, J.; Sargent, E. H. Small Alkali Cations Direct CO Electroreduction to Hydrocarbons Rather than Oxygenates. *Nat. Chem.* **2026**. https://doi.org/10.1038/s41557-025-02061-x.
(7) Shin, S.-J.; Choi, H.; Ringe, S.; Won, D. H.; Oh, H.-S.; Kim, D. H.; Lee, T.; Nam, D.-H.; Kim, H.; Choi, C. H. A Unifying Mechanism for Cation Effect Modulating C1 and C2 Productions from $CO_2$ Electroreduction. *Nat. Commun.* **2022**, *13* (1), 5482. https://doi.org/10.1038/s41467-022-33199-8.
(8) Strmcnik, D.; Kodama, K.; Van Der Vliet, D.; Greeley, J.; Stamenkovic, V. R.; Marković, N. M. The Role of Non-Covalent Interactions in Electrocatalytic Fuel-Cell Reactions on Platinum. *Nat. Chem.* **2009**, *1* (6), 466–472. https://doi.org/10.1038/nchem.330.
(9) Shao, M.; Liu, P.; Adzic, R. R. Superoxide Anion Is the Intermediate in the Oxygen Reduction Reaction on Platinum Electrodes. *J. Am. Chem. Soc.* **2006**, *128* (23), 7408–7409. https://doi.org/10.1021/ja061246s.
(10) Zhang, G.-R.; Wolker, T.; Sandbeck, D. J. S.; Munoz, M.; Mayrhofer, K. J. J.; Cherevko, S.; Etzold, B. J. M. Tuning the Electrocatalytic Performance of Ionic Liquid Modified Pt Catalysts for the Oxygen Reduction Reaction via Cationic Chain Engineering. *ACS Catal.* **2018**, *8* (9), 8244–8254. https://doi.org/10.1021/acscatal.8b02018.
(11) Angelucci, C. A.; Souza-Garcia, J.; Feliu, J. M. The Role of Adsorbates in Electrocatalytic Systems: An Analysis of Model Systems with Single Crystals. *Curr. Opin. Electrochem.* **2021**, *26*, 100666. https://doi.org/10.1016/j.coelec.2020.100666.
(12) Wei, J.; Zhang, Z.; Gee, W.; Wei, Y.; Zhou, Y.-W.; Herran, M.; Sautet, P.; Alexandrova, A. N.; Roldan Cuenya, B.; Kley, C. S. Role of Surface Hydroxyls in Atomic-Scale Copper Restructuring during CO Electroreduction. *J. Am. Chem. Soc.* **2025**, *147* (49), 45178–45188. https://doi.org/10.1021/jacs.5c14516.
(13) Sankarasubramanian, S.; Kahky, J.; Ramani, V. Tuning Anion Solvation Energetics Enhances Potassium–Oxygen Battery Performance. *Proc. Natl. Acad. Sci.* **2019**, *116* (30), 14899–14904. https://doi.org/10.1073/pnas.1901329116.
(14) Han, H.; Hemp, J.; Pace, L. A.; Ouyang, H.; Ganesan, K.; Roh, J. H.; Daldal, F.; Blanke, S. R.; Gennis, R. B. Adaptation of Aerobic Respiration to Low $O_2$ Environments. *Proc. Natl. Acad. Sci.* **2011**, *108* (34), 14109–14114. https://doi.org/10.1073/pnas.1018958108.
(15) Wang, S.; Zhu, E.; Huang, Y.; Heinz, H. Direct Correlation of Oxygen Adsorption on Platinum-Electrolyte Interfaces with the Activity in the Oxygen Reduction Reaction. *Sci. Adv.* **2021**, *7* (24), eabb1435. https://doi.org/10.1126/sciadv.abb1435.
(16) Liu, S.; Liu, S.; Bao, J.; Huang, Z.; Wei, L.; Chen, N.; Hu, Z.; Huang, W.; Pao, C.; Kong, Q.; Han, J.; Li, L.; Huang, X. Optimized Adsorption of $H_{ad}$ and $OH_{ad}$ over Amorphous $SrRuPtO_xH_y$ Nanobelts towards Efficient Alkaline Fuel Cell Catalysis. *Angew. Chem. Int. Ed.* **2025**, *64* (12), e202421013. https://doi.org/10.1002/anie.202421013.
(17) Hamada, A. T.; Orhan, M. F.; Kannan, A. M. Alkaline Fuel Cells: Status and Prospects. *Energy Rep.* **2023**, *9*, 6396–6418. https://doi.org/10.1016/j.egyr.2023.05.276.

(18) Rizo, R.; Herrero, E.; Feliu, J. M. Oxygen Reduction Reaction on Stepped Platinum Surfaces in Alkaline Media. *Phys. Chem. Chem. Phys.* **2013**, *15* (37), 15416. https://doi.org/10.1039/c3cp51642c.

(19) Kumeda, T.; Laverdure, L.; Honkala, K.; Melander, M. M.; Sakaushi, K. Cations Determine the Mechanism and Selectivity of Alkaline Oxygen Reduction Reaction on Pt(111). *Angew. Chem. Int. Ed.* **2023**, *62* (51), e202312841. https://doi.org/10.1002/anie.202312841.

(20) Yang, Y.; Agarwal, R. G.; Hutchison, P.; Rizo, R.; Soudackov, A. V.; Lu, X.; Herrero, E.; Feliu, J. M.; Hammes-Schiffer, S.; Mayer, J. M.; Abruña, H. D. Inverse Kinetic Isotope Effects in the Oxygen Reduction Reaction at Platinum Single Crystals. *Nat. Chem.* **2023**, *15* (2), 271–277. https://doi.org/10.1038/s41557-022-01084-y.

(21) Fröhlich, N. L.; Sjö, H.; Mascaró, F. V.; Koper, M. T. M. Correlating Surface Structure and Electrochemical Properties of Polycrystalline Platinum Electrodes. *Electrochim. Acta* **2026**, *548*, 147977. https://doi.org/10.1016/j.electacta.2025.147977.

(22) Li, H.; Liang, Y.; Ju, W.; Schneider, O.; Stimming, U. In Situ Monitoring of the Surface Evolution of a Silver Electrode from Polycrystalline to Well-Defined Structures. *Langmuir* **2022**, *38* (48), 14981–14987. https://doi.org/10.1021/acs.langmuir.2c02748.

(23) Larsson, A.; Grespi, A.; Vodeb, O.; Van Den Akker, K.; Ti, A.; Berschauer, C.; Imre, A. M.; Kofoed, P. M.; Lira, E.; Ramakrishnan, M.; Ansell, S.; Just, J.; Grönbeck, H.; Diebold, U.; Lundgren, E.; Merte, L. R.; Strmcnik, D.; Mom, R.; Koper, M. T. M. Platinum Surface Oxides Govern the Cathodic Overpotential of the Oxygen Reduction Reaction. *EES Catal.* **2026**, 10.1039.D6EY00014B. https://doi.org/10.1039/D6EY00014B.

(24) Herzog, A.; Lopez Luna, M.; Jeon, H. S.; Rettenmaier, C.; Grosse, P.; Bergmann, A.; Roldan Cuenya, B. Operando Raman Spectroscopy Uncovers Hydroxide and CO Species Enhance Ethanol Selectivity during Pulsed $CO_2$ Electroreduction. *Nat. Commun.* **2024**, *15* (1), 3986. https://doi.org/10.1038/s41467-024-48052-3.

(25) Chen, H.-W.; Shi, Z.-Z.; Moreno Fernandez, H.; Yang, Q.; Zheng, S.; Li, J.-T.; Li, J.-F.; Zhou, Y.; Sun, S.-G. Sensitivity and Vulnerability of the Pt Surface to Alkaline Solution. *ACS Catal.* **2026**, *16* (5), 4749–4759. https://doi.org/10.1021/acscatal.5c08275.

(26) Nayak, S.; McPherson, I. J.; Vincent, K. A. Adsorbed Intermediates in Oxygen Reduction on Platinum Nanoparticles Observed by In Situ IR Spectroscopy. *Angew. Chem.* **2018**, *130* (39), 13037–13040. https://doi.org/10.1002/ange.201804978.

(27) Nakamura, M.; Nakajima, Y.; Hoshi, N.; Tajiri, H.; Sakata, O. Effect of Non-Specifically Adsorbed Ions on the Surface Oxidation of Pt(111). *ChemPhysChem* **2013**, *14* (11), 2426–2431. https://doi.org/10.1002/cphc.201300404.

(28) Kumeda, T.; Kubo, R.; Hoshi, N.; Nakamura, M. Activation of Oxygen Reduction Reaction on Well-Defined Pt Electrocatalysts in Alkaline Media Containing Hydrophobic Organic Cations. *ACS Appl. Energy Mater.* **2019**, *2* (5), 3904–3909. https://doi.org/10.1021/acsaem.9b00582.

(29) Rizo, R.; Fernández-Vidal, J.; Hardwick, L. J.; Attard, G. A.; Vidal-Iglesias, F. J.; Climent, V.; Herrero, E.; Feliu, J. M. Investigating the Presence of Adsorbed Species on Pt Steps at Low Potentials. *Nat. Commun.* **2022**, *13* (1), 2550. https://doi.org/10.1038/s41467-022-30241-7.

(30) Wang, H.; Abruña, H. D. Identifying Adsorbed OH Species on Pt and Ru Electrodes with Surface-Enhanced Infrared Absorption Spectroscopy through CO Displacement. *J. Am. Chem. Soc.* **2023**, *145* (33), 18439–18446. https://doi.org/10.1021/jacs.3c04785.

(31) Huang, Y.-F.; Kooyman, P. J.; Koper, M. T. M. Intermediate Stages of Electrochemical Oxidation of Single-Crystalline Platinum Revealed by in Situ Raman Spectroscopy. *Nat. Commun.* **2016**, *7* (1), 12440. https://doi.org/10.1038/ncomms12440.
(32) Sun, X.; Cao, X.; Han, J.; Ji, C.; Varela, H.; Del Colle, V.; Zhang, J.; Pan, C.; Gao, Q. Effect of Electrolyte Ions on Crystalline/Amorphous α-$PtO_2$ Formation in the Electrocatalytic Oxidation of Pt(100) Preferentially Oriented Nanoparticles. *ACS Catal.* **2023**, *13* (22), 14753–14762. https://doi.org/10.1021/acscatal.3c03548.
(33) Coleman, E. J.; Co, A. C. The Complex Inhibiting Role of Surface Oxide in the Oxygen Reduction Reaction. *ACS Catal.* **2015**, *5* (12), 7299–7311. https://doi.org/10.1021/acscatal.5b02122.
(34) Jacobse, L.; Huang, Y.-F.; Koper, M. T. M.; Rost, M. J. Correlation of Surface Site Formation to Nanoisland Growth in the Electrochemical Roughening of Pt(111). *Nat. Mater.* **2018**, *17* (3), 277–282. https://doi.org/10.1038/s41563-017-0015-z.
(35) Kumeda, T.; Kondo, K.; Tanaka, S.; Sakata, O.; Hoshi, N.; Nakamura, M. Surface Extraction Process During Initial Oxidation of Pt(111): Effect of Hydrophilic/Hydrophobic Cations in Alkaline Media. *J. Am. Chem. Soc.* **2024**, *146* (15), 10312–10320. https://doi.org/10.1021/jacs.3c11334.
(36) Valls Mascaró, F.; McCrum, I. T.; Koper, M. T. M.; Rost, M. J. Nucleation and Growth of Dendritic Islands during Platinum Oxidation-Reduction Cycling. *J. Electrochem. Soc.* **2022**, *169* (11), 112506. https://doi.org/10.1149/1945-7111/ac9bdb.
(37) Sugimura, F.; Sakai, N.; Nakamura, T.; Nakamura, M.; Ikeda, K.; Sakai, T.; Hoshi, N. In Situ Observation of Pt Oxides on the Low Index Planes of Pt Using Surface Enhanced Raman Spectroscopy. *Phys. Chem. Chem. Phys.* **2017**, *19* (40), 27570–27579. https://doi.org/10.1039/C7CP04277A.
(38) Kukunuri, S.; Noguchi, H. In Situ Spectroscopy Study of Oxygen Reduction Reaction Intermediates at the Pt/Acid Interface: Surface-Enhanced Infrared Absorbance Spectroscopy. *J. Phys. Chem. C* **2020**, *124* (13), 7267–7273. https://doi.org/10.1021/acs.jpcc.9b11950.
(39) Dong, J.-C.; Zhang, X.-G.; Briega-Martos, V.; Jin, X.; Yang, J.; Chen, S.; Yang, Z.-L.; Wu, D.-Y.; Feliu, J. M.; Williams, C. T.; Tian, Z.-Q.; Li, J.-F. In Situ Raman Spectroscopic Evidence for Oxygen Reduction Reaction Intermediates at Platinum Single-Crystal Surfaces. *Nat. Energy* **2018**, *4* (1), 60–67. https://doi.org/10.1038/s41560-018-0292-z.
(40) Ze, H.; Chen, X.; Wang, X.-T.; Wang, Y.-H.; Chen, Q.-Q.; Lin, J.-S.; Zhang, Y.-J.; Zhang, X.-G.; Tian, Z.-Q.; Li, J.-F. Molecular Insight of the Critical Role of Ni in Pt-Based Nanocatalysts for Improving the Oxygen Reduction Reaction Probed Using an *In Situ* SERS Borrowing Strategy. *J. Am. Chem. Soc.* **2021**, *143* (3), 1318–1322. https://doi.org/10.1021/jacs.0c12755.
(41) Sun, Y.-L.; A, Y.-L.; Yue, M.-F.; Chen, H.-Q.; Ze, H.; Wang, Y.-H.; Dong, J.-C.; Tian, Z.-Q.; Fang, P.-P.; Li, J.-F. Exploring the Effect of Pd on the Oxygen Reduction Performance of Pt by In Situ Raman Spectroscopy. *Anal. Chem.* **2022**, *94* (11), 4779–4786. https://doi.org/10.1021/acs.analchem.1c05566.
(42) Zhong, H.-L.; Ze, H.; Zhang, X.-G.; Zhang, H.; Dong, J.-C.; Shen, T.; Zhang, Y.-J.; Sun, J.-J.; Li, J.-F. In Situ SERS Probing the Effect of Additional Metals on Pt-Based Ternary Alloys toward Improving ORR Performance. *ACS Catal.* **2023**, *13* (10), 6781–6786. https://doi.org/10.1021/acscatal.3c01317.

(43) Zhou, D.; Zheng, Y. L.; Ze, H.; Ye, X.; Cai, J.; Chen, Y.-X.; Tian, Z. Q. Why Does Pt Shell Bearing Tensile Strain Still Have Superior Activity for the Oxygen Reduction Reaction? *J. Phys. Chem. C* **2022**, *126* (42), 17913–17922. https://doi.org/10.1021/acs.jpcc.2c04720.
(44) Dong, J.-C.; Su, M.; Briega-Martos, V.; Li, L.; Le, J.-B.; Radjenovic, P.; Zhou, X.-S.; Feliu, J. M.; Tian, Z.-Q.; Li, J.-F. Direct *In Situ* Raman Spectroscopic Evidence of Oxygen Reduction Reaction Intermediates at High-Index Pt(*hkl*) Surfaces. *J. Am. Chem. Soc.* **2020**, *142* (2), 715–719. https://doi.org/10.1021/jacs.9b12803.
(45) Casalongue, H. S.; Kaya, S.; Viswanathan, V.; Miller, D. J.; Friebel, D.; Hansen, H. A.; Nørskov, J. K.; Nilsson, A.; Ogasawara, H. Direct Observation of the Oxygenated Species during Oxygen Reduction on a Platinum Fuel Cell Cathode. *Nat. Commun.* **2013**, *4* (1), 2817. https://doi.org/10.1038/ncomms3817.
(46) Jia, Q.; Caldwell, K.; Ziegelbauer, J. M.; Kongkanand, A.; Wagner, F. T.; Mukerjee, S.; Ramaker, D. E. The Role of OOH Binding Site and Pt Surface Structure on ORR Activities. *J. Electrochem. Soc.* **2014**, *161* (14), F1323–F1329. https://doi.org/10.1149/2.1071412jes.
(47) Briega-Martos, V.; Cheuquepán, W.; Feliu, J. M. Detection of Superoxide Anion Oxygen Reduction Reaction Intermediate on Pt(111) by Infrared Reflection Absorption Spectroscopy in Neutral pH Conditions. *J. Phys. Chem. Lett.* **2021**, *12* (6), 1588–1592. https://doi.org/10.1021/acs.jpclett.0c03510.
(48) Zhang, D.-Y.; Ma, Z.-F.; Wang, G.; Chen, J.; Wallace, G. C.; Liu, H.-K. Preparation of Low Loading Pt/C Catalyst by Carbon Xerogel Method for Ethanol Electrooxidation. *Catal. Lett.* **2008**, *122* (1–2), 111–114. https://doi.org/10.1007/s10562-007-9354-8.
(49) Wang, J.; Wang, Z.; Li, S.; Wang, R.; Song, Y. Surface and Interface Engineering of FePt/C Nanocatalysts for Electro-Catalytic Methanol Oxidation: Enhanced Activity and Durability. *Nanoscale* **2017**, *9* (12), 4066–4075. https://doi.org/10.1039/C6NR09122A.
(50) Banerjee, I.; Kumaran, V.; Santhanam, V. Synthesis and Characterization of Au@Pt Nanoparticles with Ultrathin Platinum Overlayers. *J. Phys. Chem. C* **2015**, *119* (11), 5982–5987. https://doi.org/10.1021/jp5113284.
(51) Huang, L.; Wei, M.; Qi, R.; Dong, C.-L.; Dang, D.; Yang, C.-C.; Xia, C.; Chen, C.; Zaman, S.; Li, F.-M.; You, B.; Xia, B. Y. An Integrated Platinum-Nanocarbon Electrocatalyst for Efficient Oxygen Reduction. *Nat. Commun.* **2022**, *13* (1), 6703. https://doi.org/10.1038/s41467-022-34444-w.
(52) Tian, R.; Xie, L.; Li, Y.; Zhang, P.; Chen, S.; Qu, Y.; Wei, Q. Evaluation of Commercial Pt/C Catalysts for Anode of Sulfur Dioxide Depolarized Electrolysis. *Mater. Sci. Eng. B* **2025**, *313*, 117943. https://doi.org/10.1016/j.mseb.2024.117943.
(53) Lee, J.; Saha, B.; Vlachos, D. G. Pt Catalysts for Efficient Aerobic Oxidation of Glucose to Glucaric Acid in Water. *Green Chem.* **2016**, *18* (13), 3815–3822. https://doi.org/10.1039/C6GC00460A.
(54) Stiles, P. L.; Dieringer, J. A.; Shah, N. C.; Van Duyne, R. P. Surface-Enhanced Raman Spectroscopy. *Annu. Rev. Anal. Chem.* **2008**, *1* (1), 601–626. https://doi.org/10.1146/annurev.anchem.1.031207.112814.
(55) Liu, J.; Zhong, H.-L.; Li, X.; Yue, M.-F.; Yang, W.-M.; You, X.; Tian, J.-H.; Wang, Y.-H.; Li, J.-F. Core-Shell Nanoparticle Enhanced Raman Spectroscopy in Situ Probing the Composition and Evolution of Interfacial Species on PtCo Surfaces. *Nano Res.* **2024**, *17* (6), 4687–4692. https://doi.org/10.1007/s12274-023-5473-9.
(56) Li, J.-F.; Yang, Z.-L.; Ren, B.; Liu, G.-K.; Fang, P.-P.; Jiang, Y.-X.; Wu, D.-Y.; Tian, Z.-Q. Surface-Enhanced Raman Spectroscopy Using Gold-Core Platinum-Shell Nanoparticle Film

Electrodes: Toward a Versatile Vibrational Strategy for Electrochemical Interfaces. *Langmuir* **2006**, *22* (25), 10372–10379. https://doi.org/10.1021/la061366d.
(57) Weber, D. J.; Dosche, C.; Oezaslan, M. Fundamental Aspects of Contamination during the Hydrogen Evolution/Oxidation Reaction in Alkaline Media. *J. Electrochem. Soc.* **2020**, *167* (2), 024506. https://doi.org/10.1149/1945-7111/ab681f.
(58) Zhu, S.; Hu, X.; Zhang, L.; Shao, M. Impacts of Perchloric Acid, Nafion, and Alkali Metal Ions on Oxygen Reduction Reaction Kinetics in Acidic and Alkaline Solutions. *J. Phys. Chem. C* **2016**, *120* (48), 27452–27461. https://doi.org/10.1021/acs.jpcc.6b09769.
(59) Sheng, W.; Zhuang, Z.; Gao, M.; Zheng, J.; Chen, J. G.; Yan, Y. Correlating Hydrogen Oxidation and Evolution Activity on Platinum at Different pH with Measured Hydrogen Binding Energy. *Nat. Commun.* **2015**, *6* (1), 5848. https://doi.org/10.1038/ncomms6848.
(60) Markovic, N.; Gasteiger, H.; Ross, P. N. Kinetics of Oxygen Reduction on Pt(hkl) Electrodes: Implications for the Crystallite Size Effect with Supported Pt Electrocatalysts. *J. Electrochem. Soc.* **1997**, *144* (5), 1591–1597. https://doi.org/10.1149/1.1837646.
(61) Yang, X.; Nash, J.; Oliveira, N.; Yan, Y.; Xu, B. Understanding the pH Dependence of Underpotential Deposited Hydrogen on Platinum. *Angew. Chem.* **2019**, *131* (49), 17882–17887. https://doi.org/10.1002/ange.201909697.
(62) Li, S.; Shi, L.; Guo, Y.; Wang, J.; Liu, D.; Zhao, S. Selective Oxygen Reduction Reaction: Mechanism Understanding, Catalyst Design and Practical Application. *Chem. Sci.* **2024**, *15* (29), 11188–11228. https://doi.org/10.1039/D4SC02853H.
(63) Murata, T.; Kotsuki, K.; Murayama, H.; Tsuji, R.; Morita, Y. Metal-Free Electrocatalysts for Oxygen Reduction Reaction Based on Trioxotriangulene. *Commun. Chem.* **2019**, *2* (1), 46. https://doi.org/10.1038/s42004-019-0149-9.
(64) Chen, W.; Xiang, Q.; Peng, T.; Song, C.; Shang, W.; Deng, T.; Wu, J. Reconsidering the Benchmarking Evaluation of Catalytic Activity in Oxygen Reduction Reaction. *iScience* **2020**, *23* (10), 101532. https://doi.org/10.1016/j.isci.2020.101532.
(65) Bard, A. J.; Faulkner, L. R. *Electrochemical Methods: Fundamentals and Applications*, 2nd ed.; Wiley: New York, 2001.
(66) Zhang, Y.; Gao, X.; Weaver, M. J. Nature of Surface Bonding on Voltammetrically Oxidized Noble Metals in Aqueous Media as Probed by Real-Time Surface-Enhanced Raman Spectroscopy. *J. Phys. Chem.* **1993**, *97* (33), 8656–8663. https://doi.org/10.1021/j100135a020.
(67) McBride, J. R.; Graham, G. W.; Peters, C. R.; Weber, W. H. Growth and Characterization of Reactively Sputtered Thin-Film Platinum Oxides. *J. Appl. Phys.* **1991**, *69* (3), 1596–1604. https://doi.org/10.1063/1.347255.
(68) Graham, G. W.; Weber, W. H.; McBride, J. R.; Peters, C. R. Raman Investigation of Simple and Complex Oxides of Platinum. *J. Raman Spectrosc.* **1991**, *22* (1), 1–9. https://doi.org/10.1002/jrs.1250220102.
(69) Bizzotto, F.; Ouhbi, H.; Fu, Y.; Wiberg, G. K. H.; Aschauer, U.; Arenz, M. Examining the Structure Sensitivity of the Oxygen Evolution Reaction on Pt Single-Crystal Electrodes: A Combined Experimental and Theoretical Study. *ChemPhysChem* **2019**, *20* (22), 3154–3162. https://doi.org/10.1002/cphc.201900193.
(70) Shinzawa, R.; Otsuka, A.; Nakamura, A. Growth of Glassy Carbon Thin Films and Its pH Sensor Applications. *SN Appl. Sci.* **2019**, *1* (2), 171. https://doi.org/10.1007/s42452-019-0181-5.

(71) Jurkiewicz, K.; Pawlyta, M.; Zygadło, D.; Chrobak, D.; Duber, S.; Wrzalik, R.; Ratuszna, A.; Burian, A. Evolution of Glassy Carbon under Heat Treatment: Correlation Structure–Mechanical Properties. *J. Mater. Sci.* **2018**, *53* (5), 3509–3523. https://doi.org/10.1007/s10853-017-1753-7.
(72) Fernández-Vidal, J.; Koper, M. T. M. The Role of Surface Science in Electrocatalysis. *ACS Catal.* **2026**, *16* (4), 2925–2934. https://doi.org/10.1021/acscatal.5c07232.
(73) Van Der Niet, M. J. T. C.; Garcia-Araez, N.; Hernández, J.; Feliu, J. M.; Koper, M. T. M. Water Dissociation on Well-Defined Platinum Surfaces: The Electrochemical Perspective. *Catal. Today* **2013**, *202*, 105–113. https://doi.org/10.1016/j.cattod.2012.04.059.
(74) Janik, M. J.; McCrum, I. T.; Koper, M. T. M. On the Presence of Surface Bound Hydroxyl Species on Polycrystalline Pt Electrodes in the "Hydrogen Potential Region" (0–0.4 V-RHE). *J. Catal.* **2018**, *367*, 332–337. https://doi.org/10.1016/j.jcat.2018.09.031.
(75) Arán-Ais, R. M.; Figueiredo, M. C.; Vidal-Iglesias, F. J.; Climent, V.; Herrero, E.; Feliu, J. M. On the Behavior of the Pt(100) and Vicinal Surfaces in Alkaline Media. *Electrochim. Acta* **2011**, *58*, 184–192. https://doi.org/10.1016/j.electacta.2011.09.029.
(76) Tanaka, H.; Sugawara, S.; Shinohara, K.; Ueno, T.; Suzuki, S.; Hoshi, N.; Nakamura, M. Infrared Reflection Absorption Spectroscopy of OH Adsorption on the Low Index Planes of Pt. *Electrocatalysis* **2015**, *6* (3), 295–299. https://doi.org/10.1007/s12678-014-0245-7.
(77) Wang, Y.; Le, J.; Li, W.; Wei, J.; Radjenovic, P. M.; Zhang, H.; Zhou, X.; Cheng, J.; Tian, Z.; Li, J. In Situ Spectroscopic Insight into the Origin of the Enhanced Performance of Bimetallic Nanocatalysts towards the Oxygen Reduction Reaction (ORR). *Angew. Chem. Int. Ed.* **2019**, *58* (45), 16062–16066. https://doi.org/10.1002/anie.201908907.
(78) Santana, J. A. DFT Calculations of the Adsorption States of $O_2$ on $OH/H_2O$-Covered Pt(111). *Electrocatalysis* **2020**, *11* (6), 612–617. https://doi.org/10.1007/s12678-020-00619-6.
(79) Wang, Y.; Wang, X.; Ze, H.; Zhang, X.; Radjenovic, P. M.; Zhang, Y.; Dong, J.; Tian, Z.; Li, J. Spectroscopic Verification of Adsorbed Hydroxy Intermediates in the Bifunctional Mechanism of the Hydrogen Oxidation Reaction. *Angew. Chem. Int. Ed.* **2021**, *60* (11), 5708–5711. https://doi.org/10.1002/anie.202015571.
(80) Adžić, R. R.; Wang, J. X. Configuration and Site of $O_2$ Adsorption on the Pt(111) Electrode Surface. *J. Phys. Chem. B* **1998**, *102* (45), 8988–8993. https://doi.org/10.1021/jp981057z.
(81) Hansen, H. A.; Viswanathan, V.; Nørskov, J. K. Unifying Kinetic and Thermodynamic Analysis of 2 $e^-$ and 4 $e^-$ Reduction of Oxygen on Metal Surfaces. *J. Phys. Chem. C* **2014**, *118* (13), 6706–6718. https://doi.org/10.1021/jp4100608.
(82) Kelly, S. R.; Kirk, C.; Chan, K.; Nørskov, J. K. Electric Field Effects in Oxygen Reduction Kinetics: Rationalizing pH Dependence at the Pt(111), Au(111), and Au(100) Electrodes. *J. Phys. Chem. C* **2020**, *124* (27), 14581–14591. https://doi.org/10.1021/acs.jpcc.0c02127.
(83) Gómez-Marín, A. Ma.; Rizo, R.; Feliu, J. M. Oxygen Reduction Reaction at Pt Single Crystals: A Critical Overview. *Catal. Sci. Technol.* **2014**, *4* (6), 1685. https://doi.org/10.1039/c3cy01049j.
(84) Kolb, M. J.; Calle-Vallejo, F.; Juurlink, L. B. F.; Koper, M. T. M. Density Functional Theory Study of Adsorption of H2O, H, O, and OH on Stepped Platinum Surfaces. *J. Chem. Phys.* **2014**, *140* (13), 134708. https://doi.org/10.1063/1.4869749.
(85) Zhuang, T.-T.; Pang, Y.; Liang, Z.-Q.; Wang, Z.; Li, Y.; Tan, C.-S.; Li, J.; Dinh, C. T.; De Luna, P.; Hsieh, P.-L.; Burdyny, T.; Li, H.-H.; Liu, M.; Wang, Y.; Li, F.; Proppe, A.; Johnston, A.; Nam, D.-H.; Wu, Z.-Y.; Zheng, Y.-R.; Ip, A. H.; Tan, H.; Chen, L.-J.; Yu, S.-H.; Kelley, S. O.; Sinton, D.; Sargent, E. H. Copper Nanocavities Confine Intermediates for Efficient

Electrosynthesis of C3 Alcohol Fuels from Carbon Monoxide. *Nat. Catal.* **2018**, *1* (12), 946–951. https://doi.org/10.1038/s41929-018-0168-4.
(86) Zhang, B.; Zhang, J.; Hua, M.; Wan, Q.; Su, Z.; Tan, X.; Liu, L.; Zhang, F.; Chen, G.; Tan, D.; Cheng, X.; Han, B.; Zheng, L.; Mo, G. Highly Electrocatalytic Ethylene Production from $CO_2$ on Nanodefective Cu Nanosheets. *J. Am. Chem. Soc.* **2020**, *142* (31), 13606–13613. https://doi.org/10.1021/jacs.0c06420.
(87) Yang, P.-P.; Zhang, X.-L.; Gao, F.-Y.; Zheng, Y.-R.; Niu, Z.-Z.; Yu, X.; Liu, R.; Wu, Z.-Z.; Qin, S.; Chi, L.-P.; Duan, Y.; Ma, T.; Zheng, X.-S.; Zhu, J.-F.; Wang, H.-J.; Gao, M.-R.; Yu, S.-H. Protecting Copper Oxidation State via Intermediate Confinement for Selective $CO_2$ Electroreduction to $C_{2+}$ Fuels. *J. Am. Chem. Soc.* **2020**, *142* (13), 6400–6408. https://doi.org/10.1021/jacs.0c01699.
(88) Ai, Q.; Wu, H.; Bonagiri, L. K. S.; Panse, K. S.; Zhou, S.; Zhao, F.; Li, Y.; Schweizer, K. S.; Aluru, N. R.; Zhang, Y. Liquid Structure Adjacent to Solid Surfaces Follows the Superposition Principle. arXiv March 27, 2026. https://doi.org/10.48550/arXiv.2603.25992.
(89) Luo, M.; Koper, M. T. M. A Kinetic Descriptor for the Electrolyte Effect on the Oxygen Reduction Kinetics on Pt(111). *Nat. Catal.* **2022**, *5* (7), 615–623. https://doi.org/10.1038/s41929-022-00810-6.
(90) Xu, P.; Wang, R.; Zhang, H.; Carnevale, V.; Borguet, E.; Suntivich, J. Cation Modifies Interfacial Water Structures on Platinum during Alkaline Hydrogen Electrocatalysis. *J. Am. Chem. Soc.* **2024**, *146* (4), 2426–2434. https://doi.org/10.1021/jacs.3c09128.
(91) Shah, A. H.; Zhang, Z.; Huang, Z.; Wang, S.; Zhong, G.; Wan, C.; Alexandrova, A. N.; Huang, Y.; Duan, X. The Role of Alkali Metal Cations and Platinum-Surface Hydroxyl in the Alkaline Hydrogen Evolution Reaction. *Nat. Catal.* **2022**, *5* (10), 923–933. https://doi.org/10.1038/s41929-022-00851-x.
(92) Murata, A.; Hori, Y. Product Selectivity Affected by Cationic Species in Electrochemical Reduction of $CO_2$ and CO at a Cu Electrode. *Bull. Chem. Soc. Jpn.* **1991**, *64* (1), 123–127. https://doi.org/10.1246/bcsj.64.123.
(93) Marcus, Y. Ionic Radii in Aqueous Solutions. *Chem. Rev.* **1988**, *88* (8), 1475–1498. https://doi.org/10.1021/cr00090a003.
(94) Govindarajan, N.; Chu, A. T.; Hahn, C.; Surendranath, Y. The Overlooked Role of Adsorption Isotherms in Electrocatalysis. *Nat. Catal.* **2025**, *8* (12), 1254–1259. https://doi.org/10.1038/s41929-025-01461-z.

**TOC Graphic**

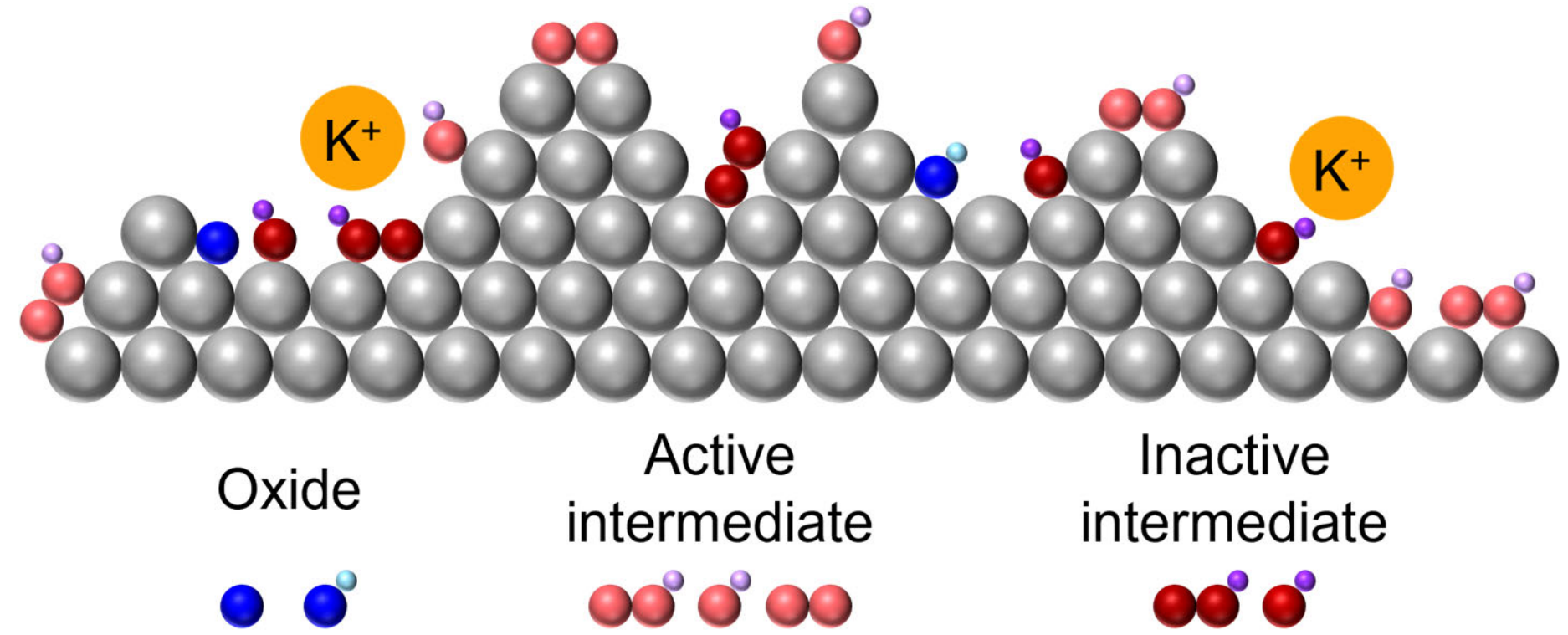